\documentclass[11pt]{article}
\usepackage{sao1}
\usepackage{psfig}

\advance\textheight by\topskip 

\def\degr{\hbox{$^\circ$}}
\def\arcmin{\hbox{$^\prime$}}

\begin{document}
\begin{center}

{\LARGE\bf Peculiar Motions in the Region of the Ursa\\
Major Supercluster of Galaxies}

\vspace{0.8cm}
{\bf Flera G. Kopylova$^*$ and Alexander I. Kopylov}

{\it Special Astrophysical Observatory, Russian Academy of Sciences,
Nizhnii Arkhyz,\\
369167 Karachai-Cherkessian Republic, Russia}

\begin{abstract}
We have investigated the peculiar motions of clusters of
galaxies in the Ursa Major (UMa) supercluster and its neighborhood.
Based on SDSS (Sloan Digital Sky Survey) data, we have compiled
a sample of early-type galaxies and used their fundamental plane to
determine the cluster distances and peculiar velocities. The samples of
early-type galaxies in the central regions (within $R_{200}$) of 12 UMa
clusters of galaxies, in three main subsystems of the supercluster --- the
filamentary structures connecting the clusters, and in nine clusters from
the nearest UMa neighborhood have similar parameters. The fairly
high overdensity (3 by the galaxy number and 15 by the cluster number)
suggests that the supercluster as a whole is gravitationally bound, while
no significant peculiar motions have been found: the peculiar velocities do
not exceed the measurement errors by more than a factor of 1.5--2. The
mean random peculiar velocities of clusters and the systematic deviations
from the overall Hubble expansion in the supercluster are consistent with
theoretical estimates. For the possible approach of the three UMa subsystems
to be confirmed, the measurement accuracy must be increased by a factor
of 2--3.
\end{abstract}
\end{center}

\renewcommand{\thefootnote}{$*$}
\footnotetext{E-mail: flera@sao.ru}

\newpage
\section{Introduction}
\hspace*{\parindent}
The superclusters of galaxies, the largest systems
in the Universe, are giant filamentary structures
in the large-scale structure (e.g., such superclusters
as Perseus (Gregory et al. 2000)
Pisces -- Cetus (Porter and Raychaudhury 2005))
and are not virialized structures. The peculiar motions
in the region of various superclusters were
investigated by many authors (by Aaronson et al.
(1989) in the Hydra -- Centaurus supercluster and by
Han and Mould (1992), Baffa et al. (1993), Hudson
et al. (1997), and Springob et al. (2003) in the
Perseus -- Pisces supercluster). Investigation of the
dynamical state of two large superclusters, Shapley
(Ettory et al. 1997) and Corona Borealis (Small et al.
1998), suggests that their central regions are at the
stage of gravitational collapse. Gramann and Suhhonenko
(2002) found (in terms of $\Lambda CDM$ models
using Nbodies) the fraction of such superclusters to
be insignificant.

We found the overdensity of the Ursa Major
(UMa) system in its central part ($11^h12^m - 12^h07^m$,
$53\degr35\arcmin - 56\degr45\arcmin$, $0.045<z<0.075$)
compared to the surrounding region ($09^h23^m - 10^h54^m$,
$12^h25^m - 13^h56^m$, $50\degr35\arcmin - 60\degr15\arcmin$,
$0.045<z<0.075$) with a size of about 150 Mpc from galaxies
with redshifts (according to NED data; these are mainly
SDSS DR4 data) to be $\approx3$ by the galaxy number and
$\sim15$ by the Abell cluster number (Abell et al. 1989)
together with additional clusters (according to NED
data) discovered in X-rays. This overdensity is high
enough for the system to be gravitationally bound, for
example, for the spherical case (D\"{u}nner et al. 2006).
For comparison, the galaxy number overdensity for
the richer Corona Borealis supercluster is 7 (Small
et al. 1998). The simulation performed in this paper
and the Hubble diagram constructed previously
(Kopylova and Kopylov 1998) show that the core
of the system may be at the stage of gravitational
collapse.

The UMa supercluster is a large flattened structure
(Figs. 1 and 2) with sizes of $40\times 15\times 125$ Mpc
in right ascension, declination, and line of sight,
respectively, a wall in the classification of large-scale
structure elements (Doroshkevich et al. 2004) that is
divided along the line of sight into three filamentary
subsystems (Figs. 1a--1c, 2). The UMa system is an
isolated one and is of interest in studying the
composition of the system and the evolution of its constituent
elements (galaxies, filaments, and clusters of
galaxies) as well as the dynamical state of the system 
as a whole, which characterizes the mass distribution 
on these scales. 

In this paper, we used photometric and spectroscopic measurements
from the SDSS DR4 catalog.
Our goal was to measure the peculiar motions of 
clusters of galaxies inside filaments, the motions of 
the filaments themselves, and clusters surrounding 
the UMa supercluster using the fundamental plane 
of the early-type galaxies. The paper is structured as 
follows. In the first part, we compile and describe a 
sample of early-type galaxies and give their distributions
in various parameters. In the second part,
we present the fundamental plane of the early-type 
galaxies for our sample and the peculiarities of using 
it to determine the distances. In the third part, we 
determine the peculiar velocities of various subsystems of UMa and its
neighborhood. In conclusion, we give our main results. Here, we used
the following cosmological parameters:
$\Omega_m=0.3$, $\Omega_{\Lambda}=0.7$, $H_0=70$~km~s$^{-1}$ Mpc$^{-1}$.

\section{The selection of early-type galaxies}
\hspace*{\parindent}
The fourth edition of the SDSS DR4 catalog 
(Adelman-McCarthy et al. 2006) allowed us to 
compile a sample of early-type galaxies in the supercluster region
($3^\circ.2\times7^\circ.9$) in the $r$ filter. The galaxies
were selected according to the following criteria: 
$fracDeV_r\geq 0.8$ (the parameter characterizes the
contribution from the de Vaucouleurs bulge to the galaxy surface
brightness profile), $\sigma > 100$~km~s$^{-1}$, $eClass\leq 0$
(the parameter characterizes the galaxy
spectrum --- there are no detectable emission lines in
the spectrum), and $r_{90}/r_{50}\geq 2.6$ (the concentration
index is equal to the ratio of the radii within which 
90\% and 50\% of the Petrosian fluxes are contained).
According to Strateva et al. (2001) and Kauffmann 
et al. (2003), precisely this index identifies early-type 
galaxies. Different authors use different values of the 
index: 2.86 (Shimasaku et al. 2001) and 2.5 (Blanton
et al. 2003). 

Thus, we compiled three samples of galaxies. The 
first sample includes early-type galaxies in UMa 
clusters within the radii $R_{200}$ that we determined
previously (Kopylova and Kopylov 2006). The second 
sample includes galaxies located in three large layers 
(filamentary subsystems) of the supercluster, except 
for the galaxies belonging to virialized cluster regions. 
The third sample represents clusters of galaxies surrounding
the supercluster and located within 80 Mpc
of the UMa center. These clusters will be studied in 
more detail in a subsequent paper. 

Table 1 gives the following cluster parameters: 
equatorial coordinates for epoch J2000.0, redshifts 
(relative to the Sun), radial velocity dispersion within $R_{200}$,
the radius $R_{200}$ within which the galaxy
density exceeds the critical density by a factor of 
200, and the number of galaxies with measured radial velocities within
$R_{200}$. The clusters A1279 and
RXCJ1010 were not involved in the radial velocity
determination, because there is only one galaxy and 
no galaxy with a measured velocity dispersion, respectively, in the
central region of the former cluster
and in the latter cluster in SDSS DR4. We took 
the data for UMa from our previous paper (Kopylova 
and Kopylov 2006) and will publish the data for the 
clusters from the UMa neighborhood in a subsequent 
paper. We selected 58 galaxies ($0.045 < z < 0.055$) in
UMa layer I; their mean $z$ is 0.051. The $z$ boundary
between layers II and III was chosen to be different 
from that shown in Fig. 2 of the above paper. It was 
displaced toward the higher $z$ by 0.01. Layer II contains 83 galaxies
($0.055 < z < 0.066$) with a mean
$z=0.061$ and layer III contains 57 galaxies ($0.066 < z < 0.075$)
with a mean $z=0.071$. Figure 3 shows the
distribution of galaxies belonging to these filaments in 
the plane of the sky of the supercluster. We selected 
153 galaxies in virialized regions of 12 UMa clusters 
and 80 galaxies in virialized regions of 9 clusters 
surrounding the supercluster. As a result, our sample 
consists of 431 early-type galaxies. 

Figure 4 presents the distributions of early-type 
galaxies from the three samples in the following parameters:
absolute magnitude, ellipticity $b/a$, logarithm of the velocity
dispersion, and concentration
index $r_{90}/r_{50}$. Analysis of the histograms suggests
that the early-type galaxies in all three samples have 
similar distributions. For example, galaxies fainter than $-21^m$
account for about 50\% of the galaxies in all three samples.
The axial ratio $b/a$ for the
most flattened E7-type elliptical galaxies is known 
to be 0.3, while most of the elliptical galaxies have 
$b/a>0.6$ (Bernardi et al. 2003). A visual examination
of the galaxies with $b/a<0.6$ (33\% in UMa, 27\% in
the surrounding clusters, and 35\% in the filaments)
shows that these are mostly S0-type galaxies. Our 
sample of early-type galaxies has magnitudes from 
$-20^m\div-23^m.25$; the magnitude $-20^m$ was taken as
the sample limit (there are few galaxies fainter than 
this limit). Table 2 presents the means and their standard
deviations for some of the parameters of galaxies
from each sample. 

\section{The fundamental plane and distance determination}
\hspace*{\parindent}
Most of the parameters of early-type galaxies determined from
photometric and spectroscopic observations are correlated. The
fundamental plane (FP) is the most convenient combination of parameters
for measuring the relative distances of clusters of galaxies
(Dressler et al. 1987; Djorgovski and Davis 1987).
This plane combines a spectroscopically measured 
parameter --- the central stellar velocity dispersion of
a galaxy and photometrically measured parameters ---
the effective radius within which the galaxy's luminosity is
half of its total luminosity and the effective surface brightness at
or (more commonly) within
the effective radius. The FP generally has an rms 
scatter of 0.08--0.09 corresponding to the accuracy
of determining the distance to an individual galaxy 
19--21\%. Depending on the number of galaxies used,
the accuracy in clusters is 3--10\%. Various studies
show that clusters with intense X-ray emission, i.e., 
the most massive virialized clusters (Gibbons et al. 
2001), have the smallest scatter on the FP. In the 
SDSS catalog, the model parameters of early-type 
galaxies were determined by fitting two-dimensional 
(bulge + disk) models to the observed profile by taking into account
the image quality and the atmospheric and galactic extinctions. We used
the derived parameters --- the equivalent (in a circular aperture)
effective radius ($r_e$) measured in arcseconds was converted to
kiloparsecs ($R_e$), the mean effective surface
brightness was calculated using the formula
$<\mu_e> = m_{deV}+2.5\log(2\pi r_e^2)-10\log(1+z)$, and
the central velocity dispersion $\sigma$ was reduced, as recommended,
to the standard circular aperture --- to calculate the peculiar
velocities of the clusters. A correction, $10\log(1+z)$,
was applied to the surface brightnesses; the K correction and the
evolutionary correction are approximately equal, but have opposite signs
(Poggianti 1997). Thus, all of the data were reduced 
to the comoving reference frame. 

The fundamental plane of the early-type galaxies 
based on SDSS data for determining the distances 
was obtained by minimizing the deviations relative 
to $R_e$ and in the $r$ filter is (Bernardi et al. 2003)
$\log R_e$ = $(1.17\pm0.04)\log\sigma+(0.30\pm0.01)<\mu_e>-8.022(\pm0.020)$
($N=8228$, the rms deviation is 0.088).
For orthogonal minimization, the coefficient of $\log\sigma$
is 1.49, the coefficient of surface brightness does not 
change, and the rms deviation is 0.094. Figure 5 
shows the FP for the early-type galaxies of our 
sample seen along the "long edge" --- the effective
radius. The line corresponds to the SDSS FP with 
the zero point for our sample. The scatter in Fig. 5 
is 0.082, which corresponds to an error in the distance 
to an individual galaxy of 19\%. The mean error in
the distances to clusters of galaxies is 6\%. The
residual deviations from the FP defined as
$\Delta_{FP} = \log R_e -1.17\log\sigma+0.30<\mu_e>+C$
(where C --- is the derived zero point of the sample, $-8.093$)
show a dependence on the galaxy magnitude and on other
parameters that correlate with the magnitude ($\log R_e$,
the concentration index, and the $g-r$ color index).
J\o rgensen et al. (1996) pointed to the existence of
such a dependence on the galaxy magnitude. Adding 
the color or the concentration index to the FP as 
the fourth parameter does not change the situation 
significantly. The magnitude dependence begins to 
manifest itself for galaxies fainter than $-20^m.75$.
For orthogonal regression, the magnitude dependence is
weaker, the scatter is larger, and a weak dependence 
on the velocity dispersion is observed. Therefore, we 
decided to use both regressions to determine the 
distances to clusters of galaxies and to average the 
peculiar velocities determined from these distances. 
The galaxies that deviated by more than $2.5\sigma$ (11)
were not involved in the distance determination (these 
are not shown in Fig. 5). 

\section{Peculiar motions in the UMa Supercluster region}
\hspace*{\parindent}
For low redshifts, the Hubble law is $cz=H_0r$ ---
the velocity of mutual recession of the galaxies $cz$
is proportional to the distance $r$, $H_0$ is the Hubble
constant. The Hubble law makes it possible to determine
the distance to a galaxy in the first approximation
by ignoring the peculiar velocity of the object.
Measuring the distance by a different method, we can
separate an additional component (along the line of
sight), the peculiar velocity of a galaxy or a cluster
of galaxies, from the directly observed radial velocity ($cz$):
$V_{cz}-H_0r$. In practice, the determined difference
between the zero point $C_{cl}$ of a cluster and the mean
zero point $C_{mn}$ is used to calculate the photometric
redshifts (distances) $z_{FP}=z_{spec}*10^{(C_{mn}-C_{cl})}$.The
peculiar velocities of clusters of galaxies in the comoving
reference frame are equal to the difference
between the spectroscopic and photometric redshifts:
$V_{p}=c(z_{spec}-z_{FP})/(1+z_{FP})$.

One might expect several types of peculiar motions
in the UMa supercluster and its neighborhood:
the motion of clusters along the filaments, the approach
of the filaments themselves, and the motion
of clusters from the nearest neighborhood of UMa
toward the supercluster if it is massive enough. We
confirmed our previously obtained result using the
photometric distances (statistically corrected for the
dependence on the  galaxy magnitude; Kopylova and
Kopylov 2001) measured using Kormendy's relation
that the UMa system on the whole obeys the Hubble
relation between $z_{spec}$ and $z_{FP}$ (Fig. 6). In this study,
the individual peculiar velocities of the clusters are,
on average, low; they do not exceed the measurement
errors by more than a factor of 1.5--2. The mean
dispersion around the FP (direct regression relative
to $R_e$) is 0.063 for the UMa clusters, 0.075 for the
surrounding clusters, and 0.066 for the clusters with
$N\geq8$ (N --- is the number of galaxies involved in the
cluster distance determination). The error in the relative
distances to the clusters of galaxies (the mean
for all clusters) is 6\%. The peculiar velocities of the
UMa clusters are presented in Table 3. In addition
to the cluster name, this table gives the number of
early-type galaxies involved in the cluster distance
determination, $z_{FP}$, the peculiar velocities determined
from direct regression and then from orthogonal regression,
and their mean value. The peculiar velocities of the three
filamentary structures as a whole
that constitute the UMa supercluster are presented 
in Table 4. The peculiar velocities found are low and 
comparable to the measurement errors. The following 
main conclusion can be reached: the UMa system 
begins to deviate from the overall Hubble expansion 
(along the line of sight). At least the signs of the peculiar
velocities of the filamentary structures suggest
that there is already a tendency for this deviation. 

The peculiar motions of clusters of galaxies differ in different
cosmological models and provide
constraints on the mass density in the Universe 
(see, e.g., Bahcall and Oh (1996) and references 
therein). For the clusters of galaxies in our sample 
(21 clusters), we determined the dispersion of the 
observed peculiar velocity distribution,
$<V^2>^{1/2}=1241\pm270$~km~s$^{-1}$, and, with a quadratic allowance
for the measurement errors, $558\pm80$~km~s$^{-1}$. For the
UMa supercluster, we obtained $<V^2>^{1/2}=1037\pm300$~km~s$^{-1}$;
including the measurement errors yields
$290\pm120$~km~s$^{-1}$. For the clusters of galaxies from
the nearest UMa neighborhood, we found
$<V^2>^{1/2}=1470\pm420$~km~s$^{-1}$ and, including the measurement
errors, $784\pm200$~km~s$^{-1}$. For the clusters of galaxies
with $\geq$ 8  members (the more galaxies are involved in
the cluster distance determination, the higher the accuracy), we found
$<V^2>^{1/2}=651\pm190$~km~s$^{-1}$ and,
including the errors, $0\pm160$. For comparison,
we determined the peculiar velocity dispersion for 
20 nearby rich clusters of galaxies from Gibbons 
et al. (2001) in a similar way. It was found to be 
$<V^2>^{1/2}=644\pm140$~km~s$^{-1}$ and, including the errors,
$469\pm80$~km~s$^{-1}$. Our estimates of the (random)
peculiar velocities for the clusters agree, within the 
error limits, with the dependences found by Sheth and 
Diaferio (2001) from computer simulations as part of 
the Virgo Consortium project at the corresponding 
mean ambient overdensity. 

On the whole, the pattern of peculiar velocities inside
UMa is consistent with the observed overdensity.
For a spherical mass concentration at a linear stage of 
collapse, the radial peculiar velocity is defined by the 
equation (Yahil 1985) $V_p = 1/3H_0r\Omega^{0.6}\delta(1+\delta)^{-1/4}$.
At $\delta=3$ and $H_0r=3000$~km~s$^{-1}$, we obtain
$V_p\simeq1000$~km~s$^{-1}$. However, since the UMa supercluster
is highly elongated along the line of sight, the peculiar 
velocity along its major axis should be lower than that 
for a spherical mass distribution. According to our 
data (Table 4), the radial velocities of the two extreme 
layers (filaments) relative to the central layer are
$300-400$~km~s$^{-1}$, which is consistent with the theoretical
estimate, given the low measurement accuracy. The 
overdensity in the layers (filaments) is higher than 
that in the supercluster as a whole (Fig. 3). Therefore, 
one might expect the peculiar velocities to be higher 
inside the filaments. This is probably observed at least 
for the two closest pairs of clusters: A1291B, A1318 
and Anon1, Anon2. The peculiar velocities for all 
these clusters are high and are directed oppositely in 
each of the pairs. 

The cluster A1291 in the ACO catalog (Abell 
et al. 1989) was determined as a cluster of richness class 1.
According to SDSS DR4 data, here
there are several peaks in the radial velocity distribution (Fig. 7)
from 14000 to 18500~km~s$^{-1}$, i.e.,
the galaxies may form a filamentary structure oriented 
along the line of sight. We identified two clusters: 
one cluster, A1291A, is associated with the largest 
peak (the velocity range 14100--16500~km~s$^{-1}$)and
includes the brightest cluster galaxy (no. 74 in Kopylova and
Kopylov (2001); the other cluster, A1291B,
has the velocity range 16500--18600~km~s$^{-1}$ and
may, in turn, consist of two subsystems. Both clusters 
have peculiar velocities directed to the observer. The 
cluster A1291B approaches A1318, which has a radial peculiar
velocity component directed to the UMa
center. A similar situation, but with a lower measurement
accuracy, is observed in the pair of clusters
Anon1 and Anon2 in the third filament. 

The peculiar velocity dispersion for the clusters 
of galaxies from the nearest UMa neighborhood and 
the scatter of galaxies on the FP are considerably 
larger than those for the supercluster. This can be 
explained in two ways. The clusters in the central 
regions are not virialized enough: half of them have 
clearly distinguishable subsystems in the plane of the 
sky (e.g., the cluster RXJ1033 has three subsystems). 
Since we investigate the central part of the cluster 
with radius $R_{200}$ (the cluster center is close to the
X-ray emission center), the relative motion of the 
subsystems inside the cluster may be responsible for 
the derived peculiar velocity (and the additional FP 
broadening). In addition, the influence of the supercluster
may be responsible for the peculiar velocities
of the clusters: for example, the clusters A1452 and 
A1507 closest to UMa show motions toward the 
supercluster (or the subsystem associated with the 
cluster A1436 (Fig. 3)). 

\section{Conclusions}
\hspace*{\parindent}
The UMa supercluster comprises (according 
to the radial velocity measurements of galaxies in 
SDSS) three large filamentary structures that are 
clearly distinguishable both in projection onto the 
plane of the sky and from radial velocity measurements. We
compiled a sample of early-type galaxies
in the central parts of 12 clusters of galaxies and in the 
filamentary structures constituting the UMa system 
as well as in the nine nearest clusters of galaxies in the 
UMa neighborhood. Using the fundamental plane of 
the early-type galaxies, we determined the distances 
and found the peculiar velocities of the clusters of 
galaxies in the comoving reference frame associated 
with the supercluster as a whole independently of the 
radial velocity measurements for galaxies. 

Our main conclusions are the following. 

(1) The early-type galaxies in the central regions 
of 12 UMa clusters, in the filamentary structures connecting the
clusters, and in 9 relatively isolated clusters from the nearest
UMa neighborhood have similar distributions in absolute magnitude $M_r$,
central velocity dispersion $\log\sigma$, concentration index
$r_{90}/r_{50}$, and axial ratio $b/a$.

(2) The peculiar velocities in the UMa system are 
low. They do not exceed the measurement errors by 
more than a factor of 1.5--2 and correspond in order of
magnitude to the observed overdensity in the central 
part of the supercluster relative to the mean density in 
the surrounding volume $\sim$150 Mpc in size (about 3
by the galaxy number and 15 by the cluster number). 

(3) The deviations from the Hubble law in the 
UMa system consisting of three filamentary structures (which are
clearly distinguishable both from
radial velocity measurements and in the plane of the 
sky) are insignificant, but they point to an approach of 
the filaments. To confirm the possible approach of the 
three UMa subsystems, the measurement accuracy 
must be increased by a factor of 2--3.

\bigskip
{\bf Acknowledgments}

The creation and distribution of the SDSS Archive 
has been funded by the Alfred P. Sloan Foundation, the Participating
Institutions, the National Aeronautics and Space Administration,
the National Science Foundation, the US Department
of Energy, the Japanese Monbukagakusho, and 
the Max Planck Society. The SDSS Web site is 
http://www.sdss.org/. 

\bigskip
{\bf References}
\begin{description}
\item[\rm ~1.] M. Aaronson, G. D. Bothun, M. E. Cornell, et al.,
Astrophys. J. {\bf338}, 654 (1989).
\item[\rm ~2.] G. O. Abell, H. G. Corwin, Jr., and R. P. Olowin,
Astrophys. J., Suppl. Ser. {\bf70}, 1 (1989).
\item[\rm ~3.] J. K. Adelman-McCarthy, M.A.Aqueros, S.S. Alam,
et al., Astrophys. J., Suppl. Ser. {\bf162}, 38 (2006).
\item[\rm ~4.] C. Baffa, G. Chincarini, R.B.C.Henry, and
J. Manousoyanaki, Astron. Astrophys. {\bf280}, 20 (1993).
\item[\rm ~5.] N. A. Bahcall and S. P. Oh, Astrophys. J. {\bf462},
L49 (1996).
\item[\rm ~6.] M. Bernardi, R.K.Sheth, J.Annis, et al., Astron.J.
{\bf125}, 1866 (2003).
\item[\rm ~7.] M. R. Blanton, D. W. Hogg, N. A. Bahcall, et al.,
Astrophys. J. {\bf594}, 186 (2003).
\item[\rm ~8.] S. Djorgovski and M. Davis, Astrophys. J. {\bf313}, 59 (1987).
\item[\rm ~9.] A. Doroshkevich, D.L.Tucker, S. Allam, and
M. J. Way, Astron. Astrophys. {\bf418}, 7 (2004).
\item[\rm 10.] A. Dressler, D. Lynden-Bell, D. Burstein, et al.,
Astrophys. J. {\bf313}, 42 (1987).
\item[\rm 11.] R. D\"{u}nner, P. A. Araya, A. Meza, and A. Reizenegger,
astro-ph/0603709 (2006).
\item[\rm 12.] S. Ettory, A. C. Fabian, and D. A. White, Mon. Not.
R. Astron. Soc. {\bf289}, 787 (1997).
\item[\rm 13.] R. A. Gibbons, A. S. Fruchter, and G. D. Bothun,
Astron. J. {\bf121}, 649 (2001).
\item[\rm 14.] M. Gramann and I. Suhhonenko, Mon. Not. R. Astron. Soc.
{\bf329}, L47 (2002).
\item[\rm 15.] S. A. Gregory, W. G. Tifft, J. W. Moody, et al., Astron.
J. {\bf119}, 567 (2000).
\item[\rm 16.] M. Han and J. R. Mould, Astrophys. J. {\bf396}, 453
(1992).
\item[\rm 17.] M. J. Hudson,J.R. Lucey,R. J.Smith,etal., Mon.
Not. R. Astron. Soc. {\bf291}, 488 (1997).
\item[\rm 18.] I.J\o rgensen, M. Franx, and P. Kjaergaard, Mon. Not.
R. Astron. Soc. {\bf280}, 167 (1996).
\item[\rm 19.] G. Kauffmann, T. M. Heckman, S. D. M. White, et al.,
Mon. Not. R. Astron. Soc. {\bf341}, 54 (2003).
\item[\rm 20.] F. G. Kopylova and A. I. Kopylov, Pis'ma Astron. Zh.
{\bf24}, 573 (1998) [Astron. Lett. {\bf24}, 491 (1998)].
\item[\rm 21.] F. G. Kopylova and A. I. Kopylov, Pis'ma Astron. Zh.
{\bf27}, 403 (2001) [Astron. Lett. {\bf27}, 345 (2001)].
\item[\rm 22.] F. G. Kopylova and A. I. Kopylov, Pis'ma Astron. Zh.
{\bf32}, 95 (2006) [Astron. Lett. {\bf32}, 84 (2006)].
\item[\rm 23.] B. M. Poggianti, Astron. Astrophys., Suppl. Ser. {\bf122},
399 (1997).
\item[\rm 24.] S. C. Porter and S. Raychaudhury, Mon. Not. R.
Astron. Soc. {\bf364}, 1387 (2005).
\item[\rm 25.] R. K. Sheth and A. Diaferio, Mon. Not. R. Astron.
Soc. {\bf322}, 901 (2001).
\item[\rm 26.] K. Shimasaku, M. Fukugita, M. Doi, et al., Astron. J.
{\bf122}, 1238 (2001).
\item[\rm 27.] T. A. Small, Ching-Pei Ma, W. L. W. Sargent, and
D. Hamilton, Astrophys. J. {\bf492}, 45 (1998).
\item[\rm 28.] C. M. Springob, M. P. Haynes, and R. Giovanelli,
Bull.Am. Astron.Soc. {\bf35}, 1283 (2003).
\item[\rm 29.] I. Strateva, I. \^{Z}eliko, G.R.Knapp, et al., Astron.J.
{\bf122}, 1861 (2001).
\item[\rm 30.] A. Yahil, The Virgo Cluster of Galaxies, Ed. by
O. Richter and B. Binggeli (ESO, Garching, 1985), p.359.
\end{description}

{\it{Translated by V. Astakhov}}

\newpage
\begin{center}

 Table 1. Parameters of galaxy clusters in the UMa supercluster
	  and its neighborhood

\vspace{0.4cm}
\begin{tabular}{lcccrcc}
\hline
Cluster& RA J2000.0& DEC J2000.0& $z_{spec}$& $N$& $\sigma$, km s$^{-1}$& $R_{200}$, Mpc\\
\hline
A1270   & 11 29 42.0 & +54 05 56 & 0.06890 & 53& 553& 1.24\\
A1291A  & 11 32 21.1 & +55 58 03 & 0.05092 & 17& 424& 0.97\\
A1291B  & 11 32 02.4 & +56 04 12 & 0.05715 & 10& 450& 1.02\\
A1318   & 11 36 03.5 & +55 04 31 & 0.05647 & 37& 411& 0.94\\
A1377   & 11 47 21.3 & +55 43 49 & 0.05170 & 74& 613& 1.40\\
A1383   & 11 48 05.8 & +54 38 47 & 0.05979 & 55& 527& 1.20\\
A1436   & 12 00 08.8 & +56 10 52 & 0.06517 & 71& 701& 1.60\\
Anon1   & 11 15 23.8 & +54 26 39 & 0.06944 & 41& 561& 1.25\\
Anon2   & 11 19 46.0 & +54 28 02 & 0.07056 & 13& 253& 0.60\\
Anon3   & 11 29 32.3 & +55 25 20 & 0.06806 & 23& 362& 0.81\\
Anon4   & 11 39 08.5 & +55 39 52 & 0.06118 & 24& 391& 0.89\\
Sh166   & 12 03 11.9 & +54 50 50 & 0.04996 & 18& 318& 0.73\\
 \hline
A1003   & 10 25 01.6 & +47 50 30 & 0.06282 & 29& 606& 1.37\\
A1169   & 11 07 49.3 & +43 55 00 & 0.05882 & 52& 581& 1.32\\
A1279   & 11 31 39.3 & +67 13 27 & 0.05432 &  6& 187& 0.43\\
A1452   & 12 03 28.4 & +51 42 56 & 0.06186 & 20& 513& 1.16\\
A1461   & 12 04 24.7 & +42 33 43 & 0.05405 & 12& 317& 0.72\\
A1507   & 12 14 48.6 & +59 54 22 & 0.06002 & 34& 401& 0.91\\
A1534   & 12 24 42.8 & +61 28 15 & 0.06999 & 17& 304& 0.68\\
RXCJ1010& 10 10 16.1 & +54 30 06 & 0.04581 & 44& 407& 0.94\\
RXJ1033 & 10 33 51.2 & +57 03 21 & 0.04598 & 25& 398& 0.92\\
RXCJ1053& 10 54 11.2 & +54 50 18 & 0.07186 & 47& 506& 1.13\\
RXCJ1122& 11 22 15.4 & +67 13 19 & 0.05511 & 11& 236& 0.54\\
\hline
\end{tabular}
\end{center}

\vspace{1.2cm}
\begin{center}
Table 2. Mean parameters of early-type galaxies: (1) UMa clusters,
	 (2) filaments, \\and (3) surrounding clusters

\vspace{0.4cm}
\begin{tabular}{lrccccl} \hline
Sample & $N$& $<M_r>$& $<b/a>$& $<r_{90}/r_{50}>$& $<\log\sigma>$& $<\mu_e>$\\
\hline
(1) & 153& $-21.21~(0.65)$& 0.68~(0.19)& 2.99~(0.24)& 2.24~(0.09)& $19.99~(0.55)$\\
(2) & 198& $-21.12~(0.67)$& 0.67~(0.20)& 3.02~(0.22)& 2.24~(0.09)& $20.04~(0.60)$\\
(3) &  80& $-21.04~(0.68)$& 0.70~(0.19)& 2.96~(0.24)& 2.23~(0.10)& $20.01~(0.59)$\\
\hline
\end{tabular}
\end{center}

\newpage
\begin{center}

Table 3. Peculiar velocities of clusters

\vspace{0.4cm}
\begin{tabular}{lrcrrr} \hline
Cluster& $N$& $z_{FP}$& $V_{p}(1)$, km s$^{-1}$& $V_{p}(2)$, km s$^{-1}$& $<V_{p}>$, km s$^{-1}$\\
\hline
A1270   & 18& 0.06819& $ +179^{+958}_{-1000}$ & $  +45^{+1043}_{-1093}$& $ +112^{+1000}_{-1050}$\\
A1291A  &  8& 0.05265& $ -684^{+558}_{-579}$  & $ -441^{+623}_{-648}$  & $ -562^{+590}_{-610}$  \\
A1291B  &  5& 0.06089& $-1394^{+478}_{-492}$  & $-1274^{+552}_{-568}$  & $-1334^{+520}_{-530}$  \\
A1318   &  8& 0.05288& $+1073^{+829}_{-869}$  & $ +829^{+687}_{-715}$  & $ +951^{+760}_{-790}$  \\
A1377   & 20& 0.05158& $  -68^{+468}_{-482}$  & $   +0^{+533}_{-552}$  & $  -34^{+500}_{-520}$  \\
A1383   & 22& 0.06069& $ -196^{+575}_{-593}$  & $ -474^{+575}_{-593}$  & $ -335^{+575}_{-593}$  \\
A1436   & 34& 0.06450& $ +168^{+459}_{-470}$  & $  +42^{+500}_{-514}$  & $ +105^{+480}_{-490}$  \\
Anon1   & 18& 0.07424& $-1385^{+619}_{-636}$  & $-1480^{+836}_{-867}$  & $-1432^{+730}_{-750}$  \\
Anon2   &  5& 0.07002& $+2612^{+1370}_{-1450}$& $+2081^{+1410}_{-1500}$& $+2346^{+1390}_{-1480}$\\
Anon3   &  5& 0.06791& $ +132^{+1731}_{-1879}$& $ -221^{+1527}_{-1642}$& $  -44^{+1630}_{-1760}$\\
Anon4   &  4& 0.05677& $ +587^{+1307}_{-1401}$& $ +276^{+1601}_{-1745}$& $ +432^{+1450}_{-1570}$\\
Sh166   &  6& 0.05472& $-1405^{+832}_{-910}$  & $-1441^{+1109}_{-1192}$& $-1423^{+970}_{-1050}$ \\
 \hline
A1003   &  9& 0.06421& $ +283^{+364}_{-370}$  & $  -82^{+722}_{-748}$  & $ +100^{+540}_{-560}$  \\
A1169   & 14& 0.05592& $ +641^{+751}_{-782}$  & $+1007^{+861}_{-903}$  & $ +824^{+810}_{-840}$  \\
A1452   &  6& 0.05285& $+2271^{+1729}_{-1894}$& $+2864^{+2092}_{-2340}$& $+2568^{+1910}_{-2120}$\\
A1461   &  4& 0.05177& $ +591^{+1228}_{-1323}$& $ +693^{+1228}_{-1323}$& $ +642^{+1230}_{-1320}$\\
A1507   & 15& 0.05673& $ +841^{+915}_{-961}$  & $+1026^{+915}_{-961}$  & $ +934^{+920}_{-960}$  \\
A1534   &  8& 0.06959& $ +180^{+1652}_{-1782}$& $  +46^{+1273}_{-1348}$& $ +113^{+1460}_{-1570}$\\
RXJ1033 &  4& 0.05484& $-2182^{+1494}_{-1666}$& $-2859^{+996}_{-1070}$ & $-2520^{+1240}_{-1370}$\\
RXCJ1053& 15& 0.07303& $ -561^{+820}_{-848}$  & $  -92^{+952}_{-992}$  & $ -326^{+890}_{-920}$  \\
RXCJ1122&  5& 0.06256& $-1819^{+1217}_{-1309}$& $-2383^{+1184}_{-1270}$& $-2101^{+1200}_{-1290}$\\
\hline
\end{tabular}
\end{center}

\vspace{1.2cm}
\begin{center}

Table 4.  Peculiar velocities of filaments

\vspace{0.4cm}
\begin{tabular}{crccrrr} \hline
Filament& $N$& $z_{spec}$& $z_{FP}$& $V_{p}(1)$, km s$^{-1}$& $V_{p}(2)$, km s$^{-1}$& $<V_{p}>$, km s$^{-1}$\\
\hline
I   & 58& 0.05078& 0.04906& $+395^{+330}_{-340}$& $+590^{+360}_{-370}$& $+492^{+340}_{-360}$\\
II  & 83& 0.06107& 0.06037& $+275^{+350}_{-360}$& $+119^{+350}_{-360}$& $+197^{+350}_{-360}$\\
III & 57& 0.07098& 0.07181& $ -46^{+500}_{-510}$& $-416^{+540}_{-555}$& $-231^{+520}_{-530}$\\
\hline
\end{tabular}
\end{center}

\newpage
\begin{figure}[*]
\centerline{\psfig{figure=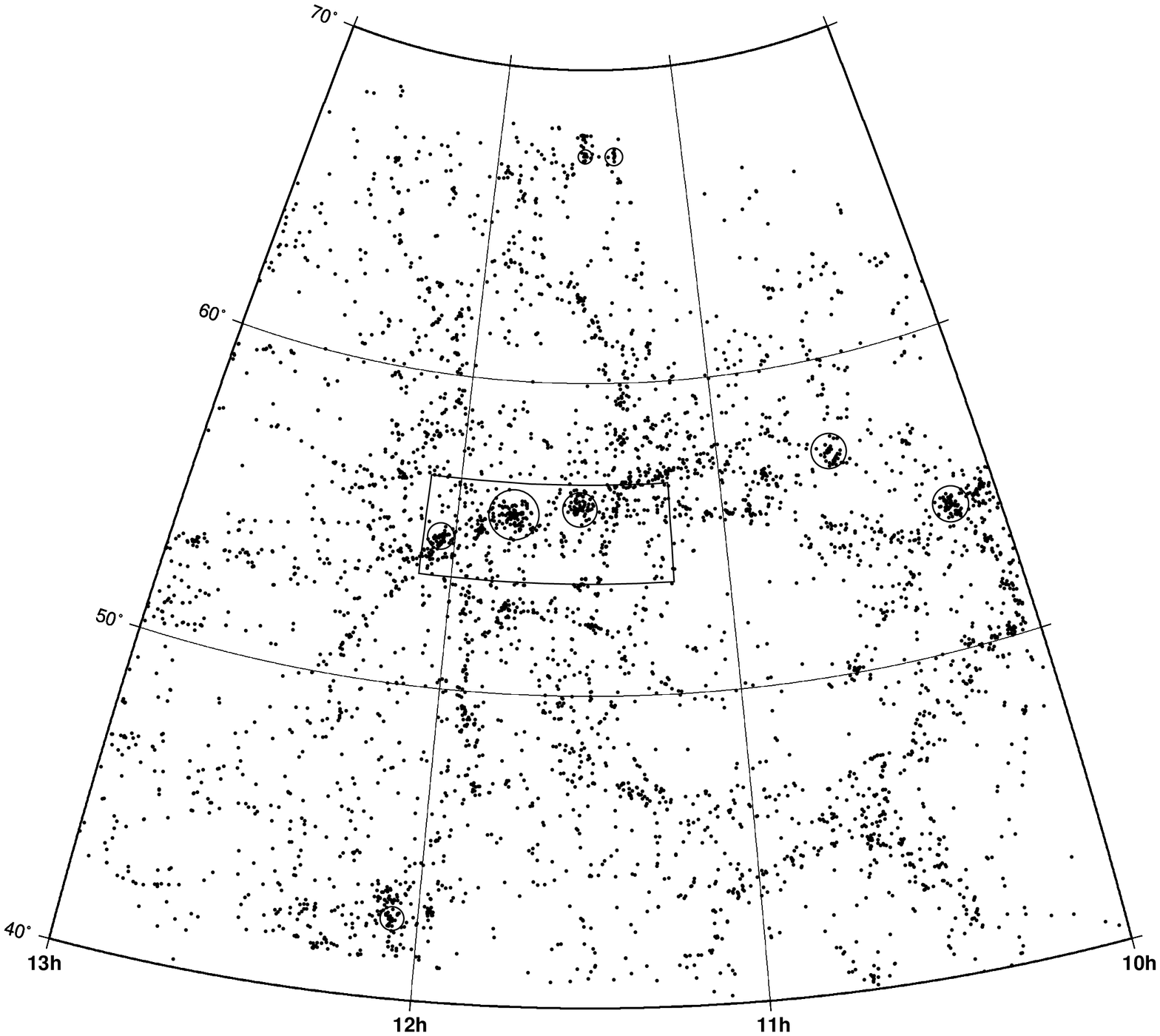,width=23cm,angle=-90,%
clip=}}
\caption[]{
Distribution of galaxies (dots) in the plane of the sky
in the redshift ranges (a) 0.045--0.055, (b) 0.055--0.065, and
(c) 0.065--0.075. The clusters of galaxies are denoted by circles
with radii of 2$R_{200}$. The central part of the UMa superclusters
is highlighted by the rectangle.
}
\end{figure}

\newpage
\begin{figure}[*]
\setcounter{figure}{0}
\centerline{\psfig{figure=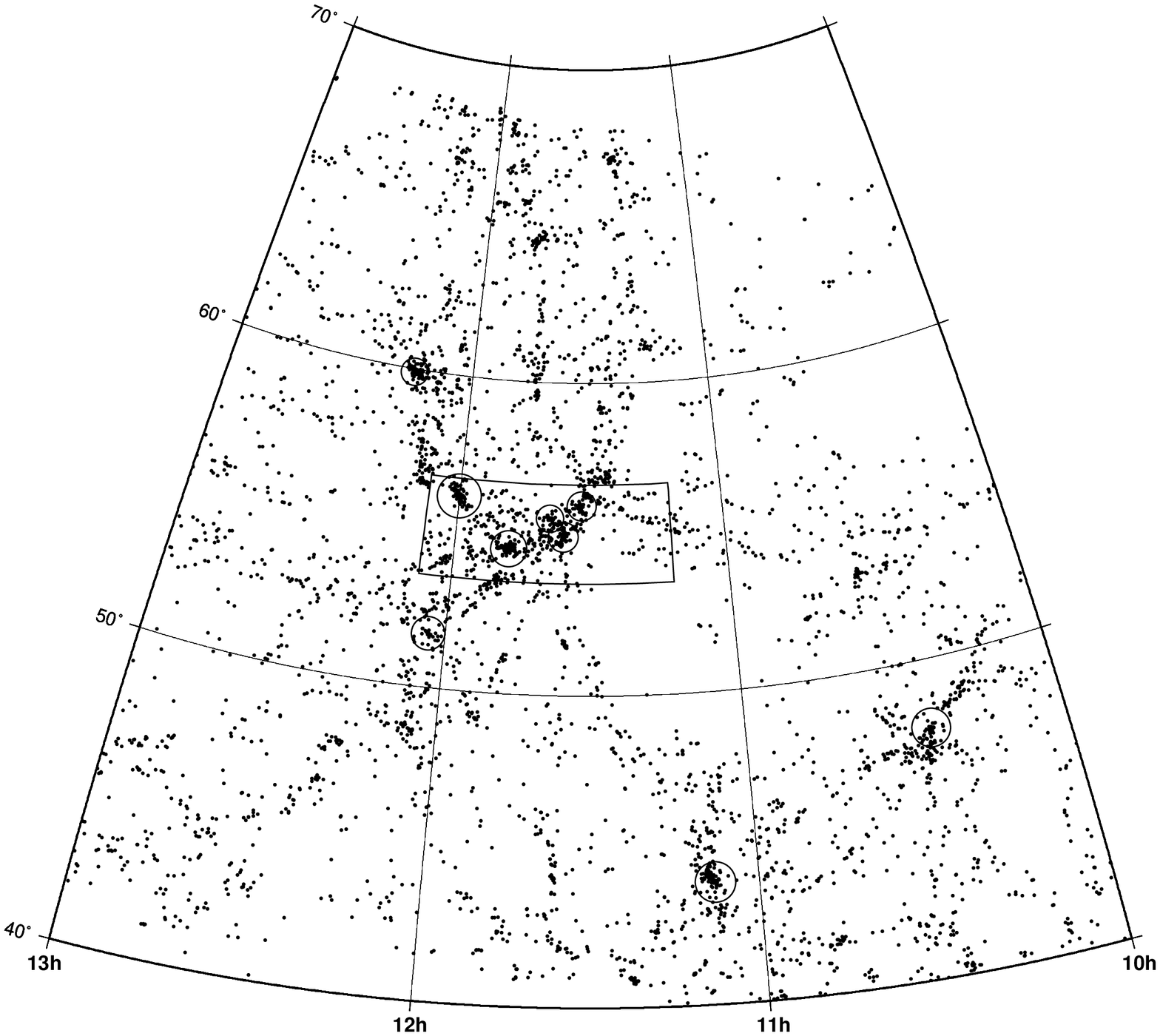,width=23cm,angle=-90,%
clip=}}
\caption[]{(b)
}
\end{figure}

\newpage
\begin{figure}[*]
\setcounter{figure}{0}
\centerline{\psfig{figure=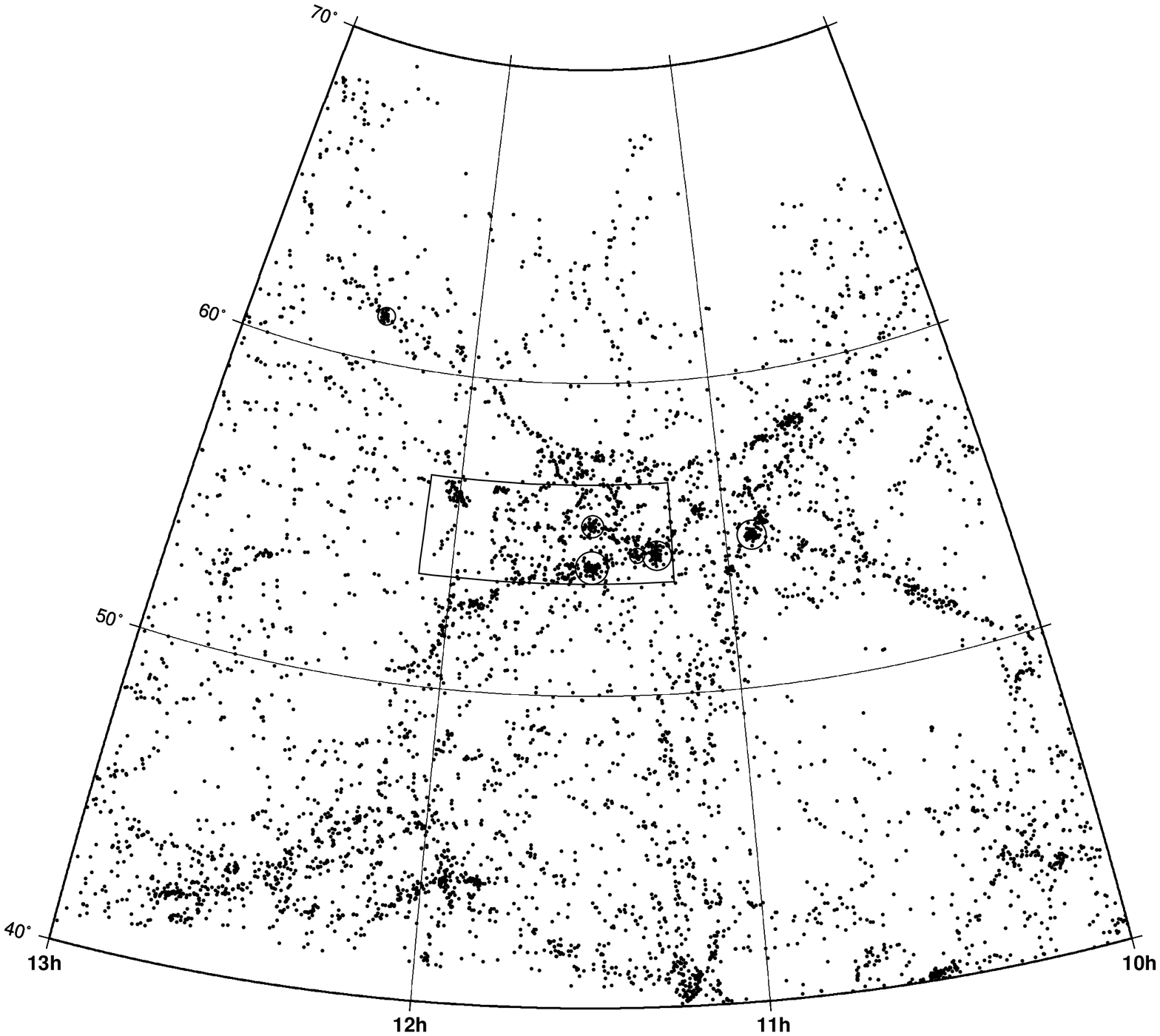,width=23cm,angle=-90,%
clip=}}
\caption[]{(c)
}
\end{figure}

\newpage
\begin{figure}[*]
\centerline{\psfig{figure=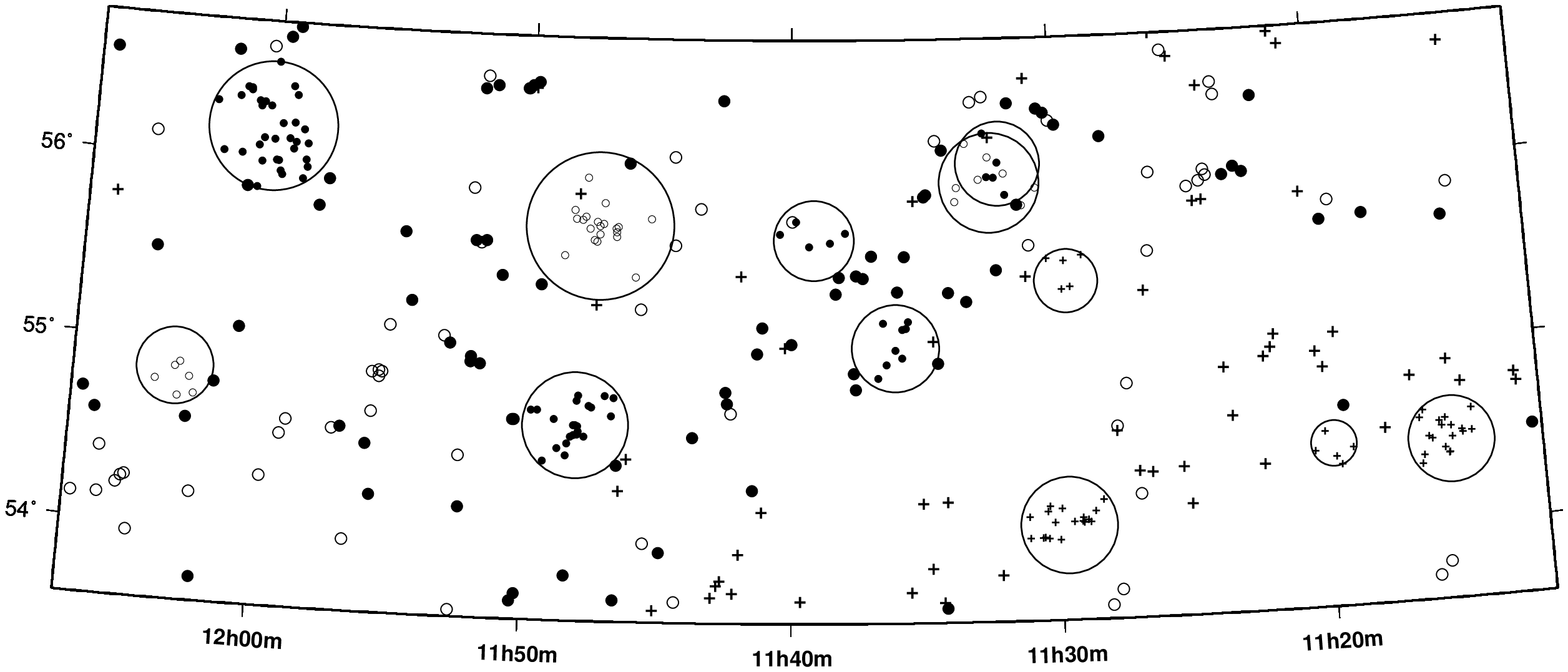,width=19cm,angle=-90,%
clip=}}
\caption[]{
Central part of the UMa supercluster. The early-type galaxies
divided into three subsystems in redshift are shown :
(I) $0.045<z<0.055$ (open circles), (II) $0.056<z<0.066$ (filled circles),
and (III) $0.066<z<0.075$ (pluses). The cluster members
are indicated by the smaller symbols in accordance with the mean
cluster redshift. The clusters are denoted by the large circles with
radii of $R_{200}$.
}
\end{figure}

\newpage
\begin{figure}[*]
\centerline{\psfig{figure=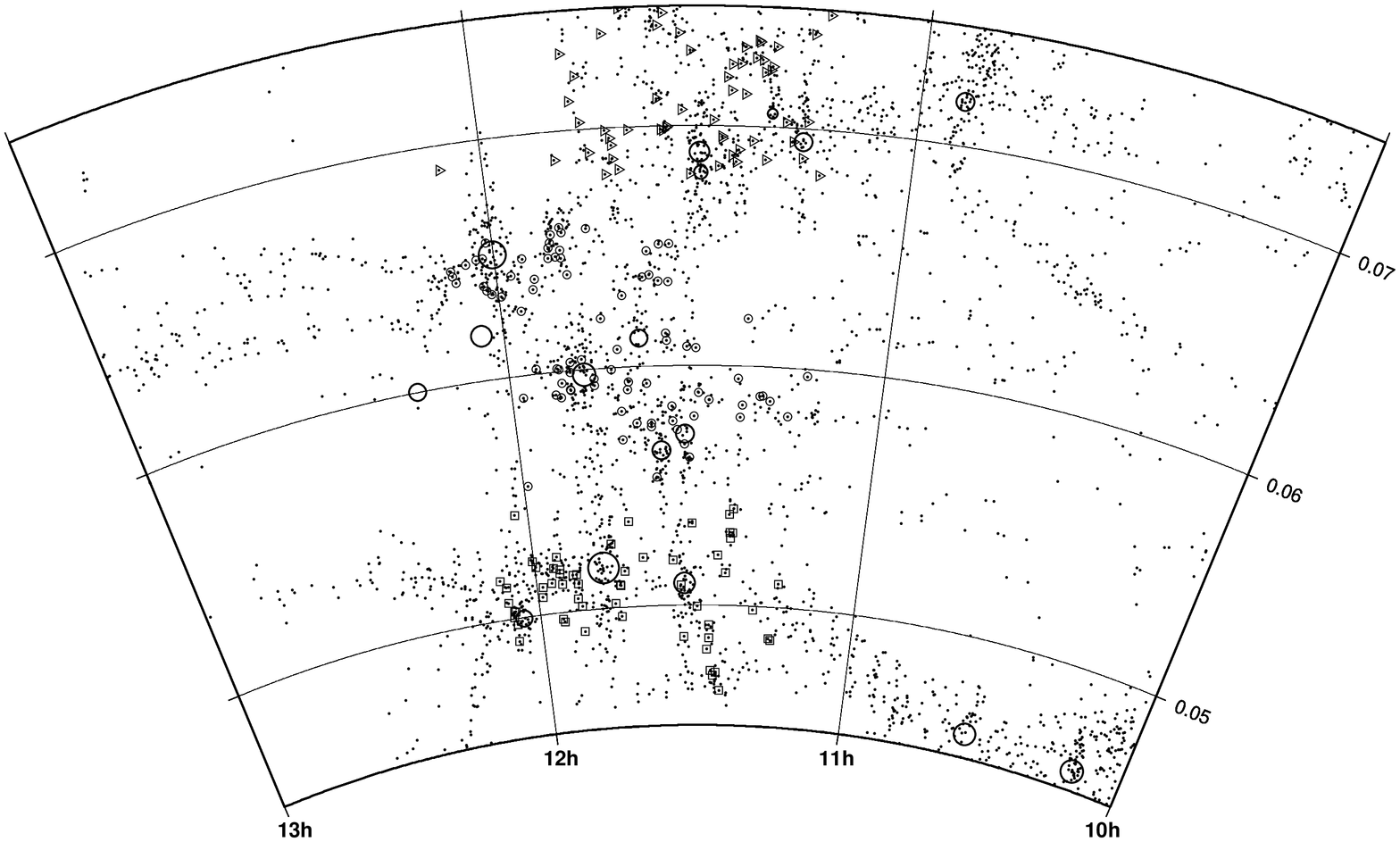,width=19cm,angle=-90,%
clip=}}
\caption[]{
Distribution of galaxies in the central part of the supercluster
and its neighborhood in the declination layer
$53\degr35\arcmin-56\degr45\arcmin$. The clusters of galaxies (in
the layer $50\degr-60\degr$) are indicated by the large
circles with radii proportional to $R_{200}$. The early-type
galaxies belonging to the three UMa subsystems are highlighted by
the squares (I), circles (II), and triangles (III).
}
\end{figure}

\newpage
\begin{figure}[*]
\centerline{\psfig{figure=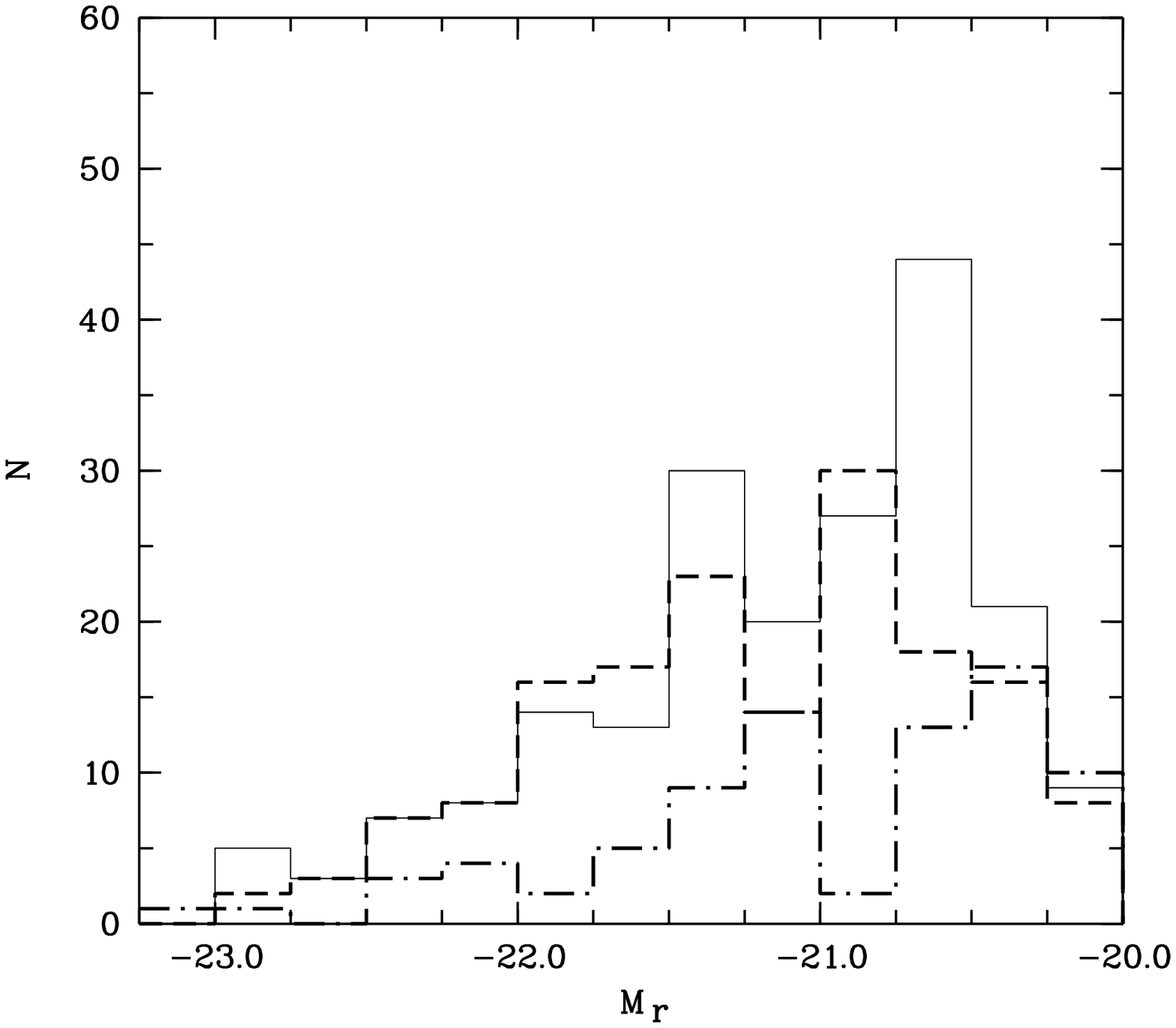,width=23cm,angle=-90,%
clip=}}
\caption[]{
Distributions of early-type galaxies in the UMa clusters ($R_{200}$)
(dashed lines), in the clusters surrounding UMa ($R_{200}$)
(dash-dotted lines), and in the filaments between the UMa
clusters (solid lines) in (a) absolute magnitude $M_r$, (b) axial ratio
(b/a), (c) logarithm of the velocity dispersion ($log\sigma$), and
(d) concentration index ($r_{90}/r_{50}$).
}
\end{figure}

\newpage
\begin{figure}[*]
\setcounter{figure}{3}
\centerline{\psfig{figure=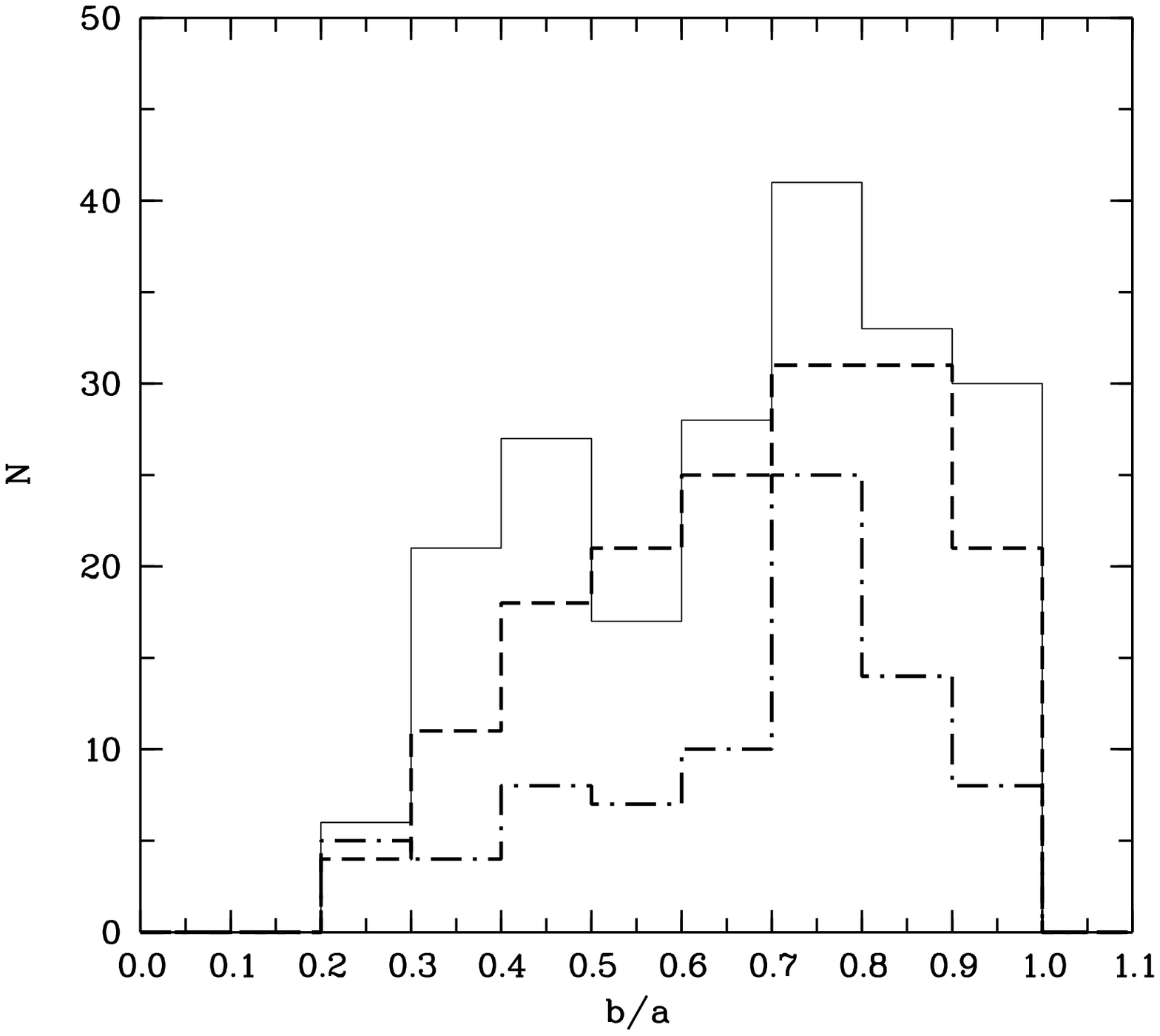,width=23cm,angle=-90,%
clip=}}
\caption[]{(b)
}
\end{figure}

\newpage
\begin{figure}[*]
\setcounter{figure}{3}
\centerline{\psfig{figure=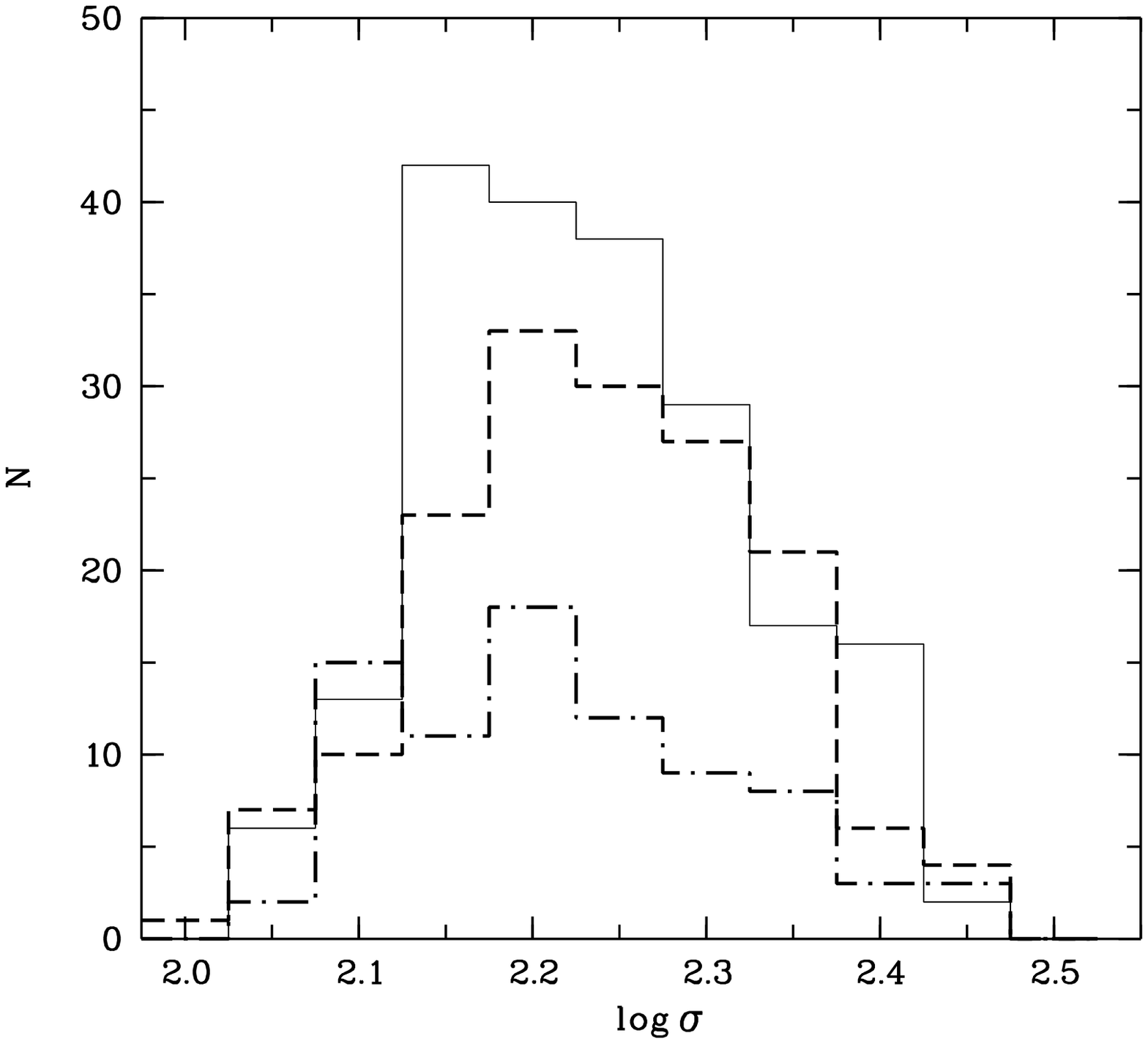,width=23cm,angle=-90,%
clip=}}
\caption[]{(c)
}
\end{figure}

\newpage
\begin{figure}[*]
\setcounter{figure}{3}
\centerline{\psfig{figure=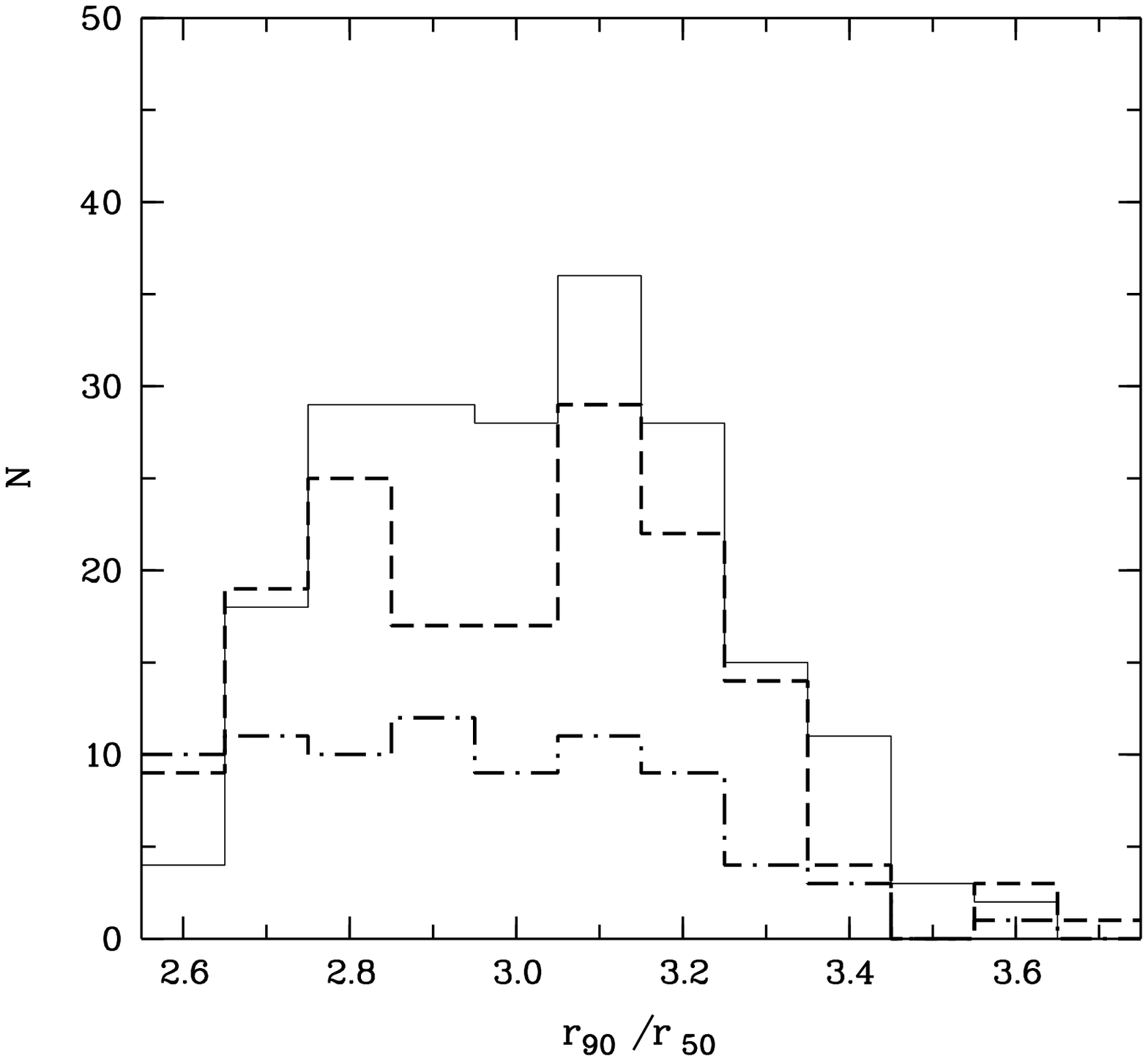,width=23cm,angle=-90,%
clip=}}
\caption[]{(d)
}
\end{figure}

\newpage
\begin{figure}[*]
\centerline{\psfig{figure=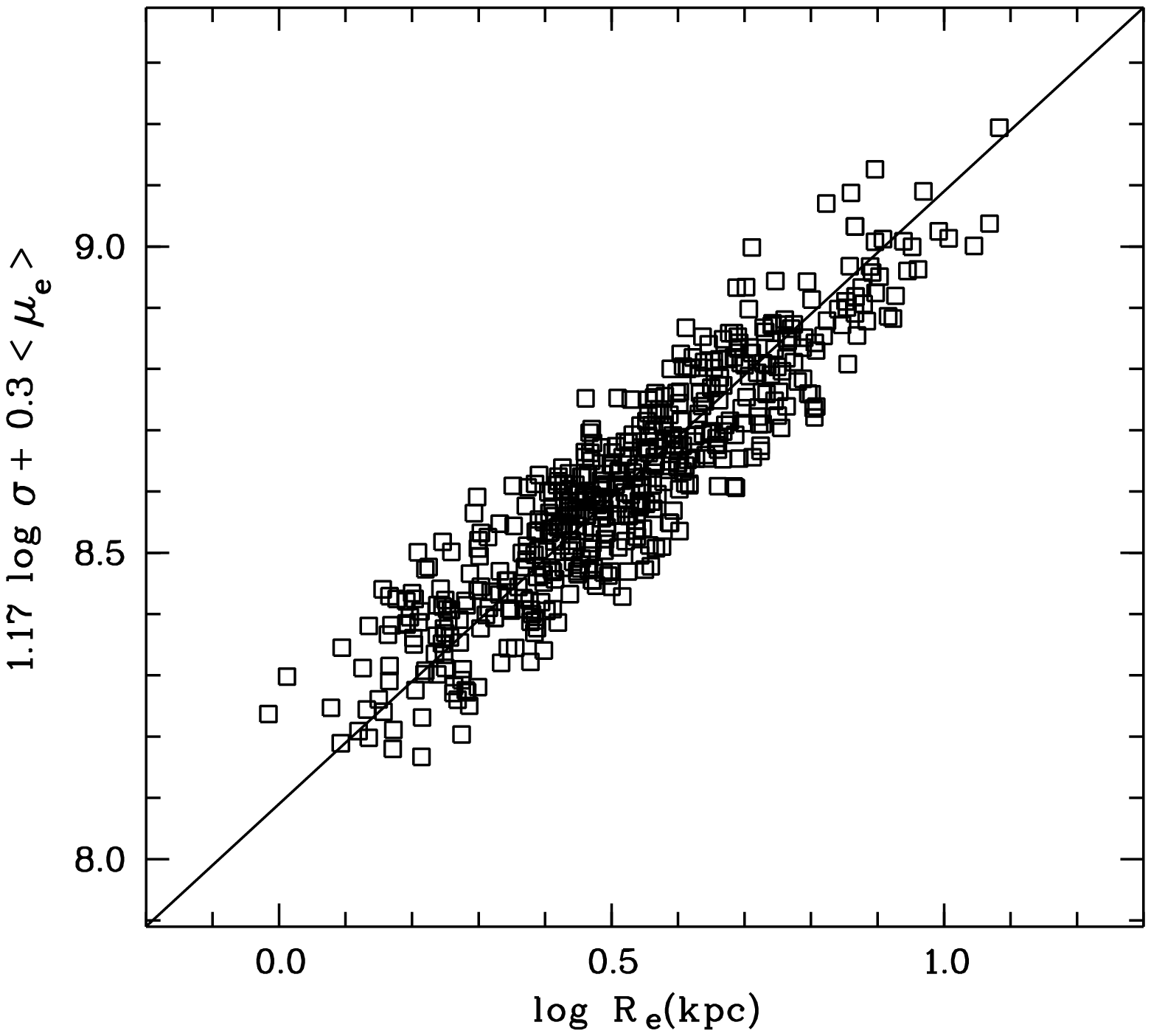,width=23cm,angle=-90,%
clip=}}
\caption[]{
Fundamental plane of the early-type galaxies. The
line corresponds to the FP formula (direct regression) and
the zero point of the sample.
}
\end{figure}

\newpage
\begin{figure}[*]
\centerline{\psfig{figure=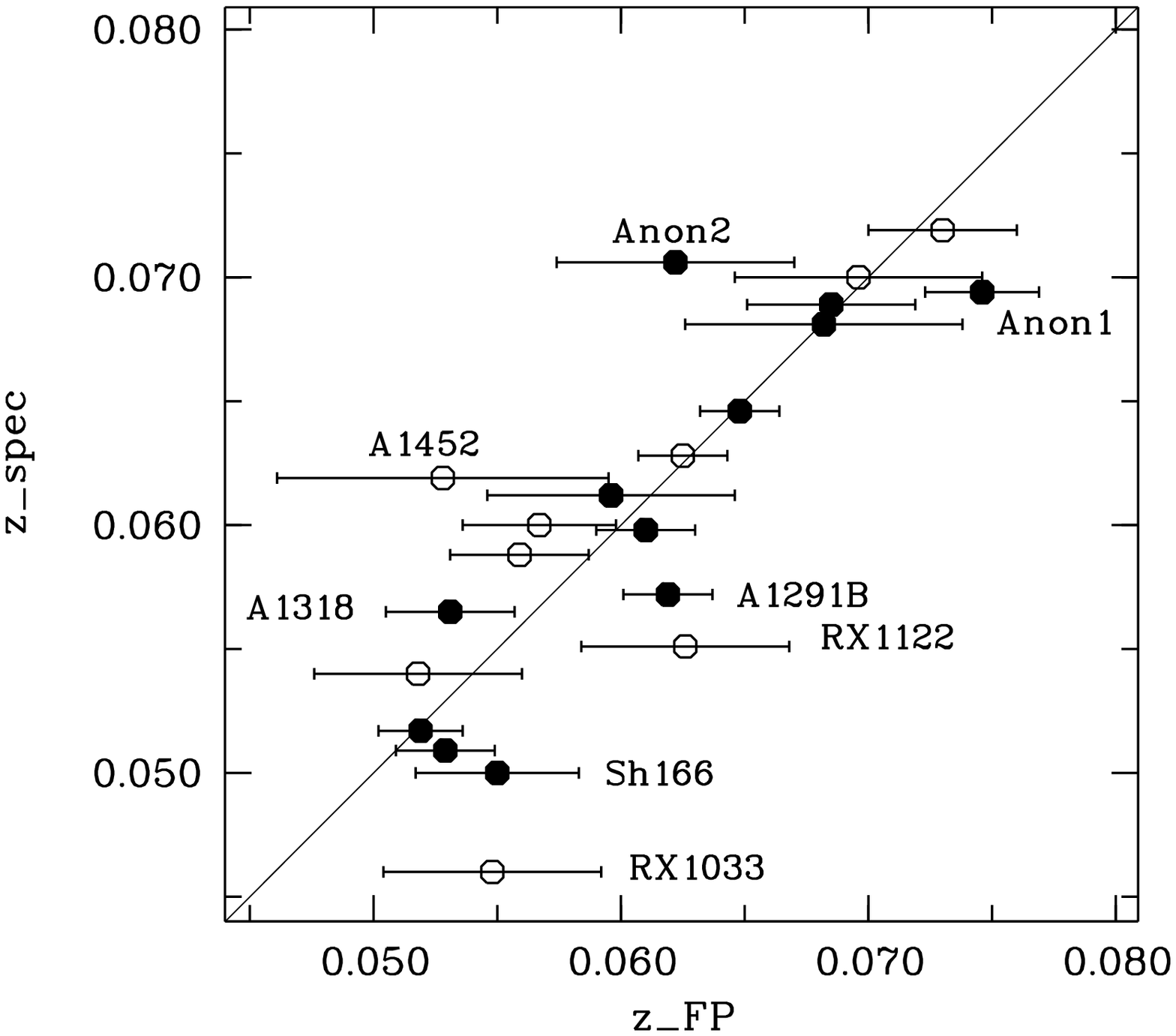,width=23cm,angle=-90,%
clip=}}
\caption[]{
Hubble diagram for the UMa supercluster (filled
circles). The clusters surrounding UMa are marked by
the open circles. The rms errors of the distances are indicated.
The names of the clusters with peculiar velocities
exceeding $1\sigma$ are given.
}
\end{figure}

\begin{figure}[*]
\centerline{\psfig{figure=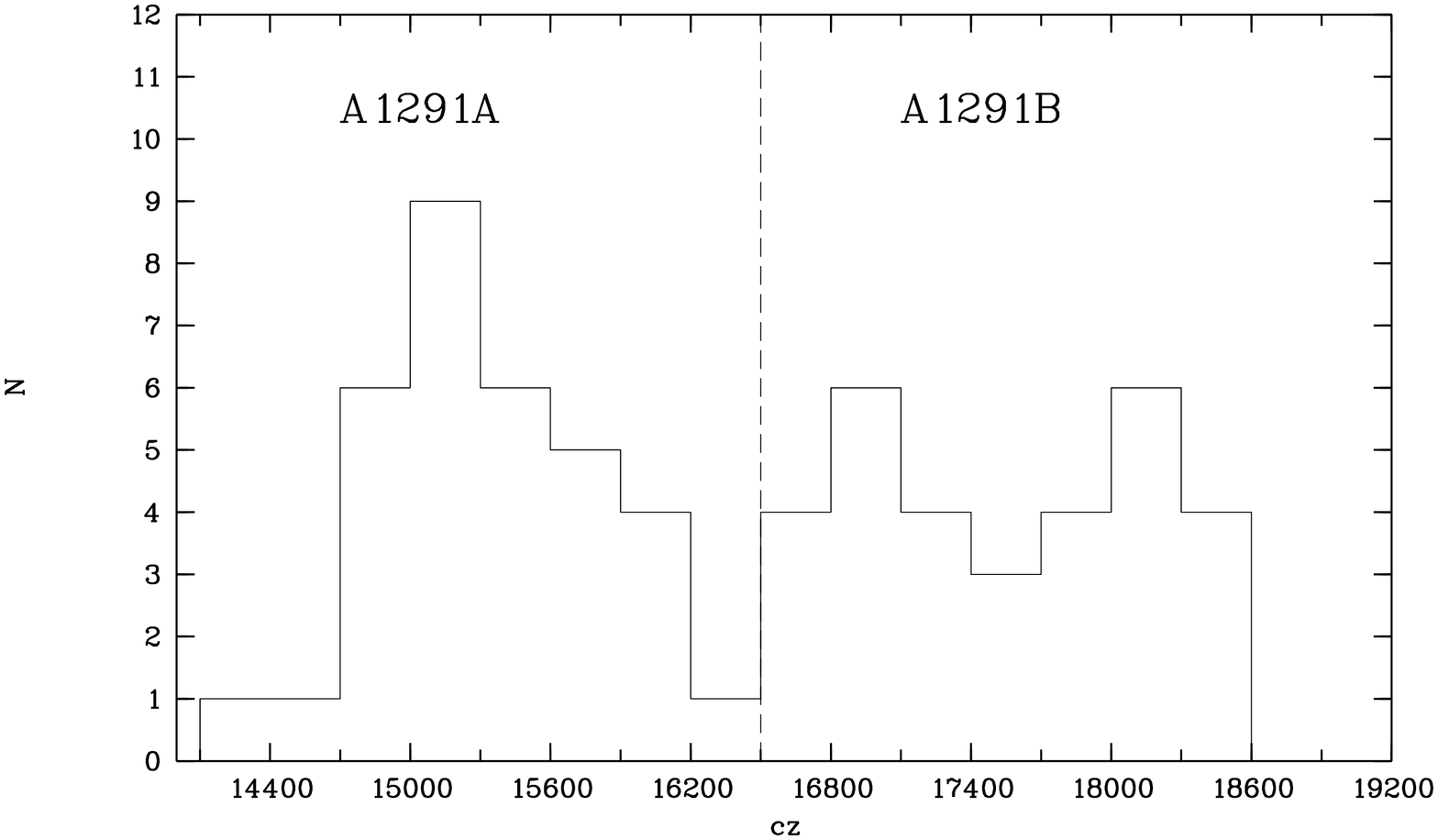,width=16cm,angle=-90,%
clip=}}
\caption[]{
Radial velocity distribution of galaxies toward the cluster A1291.
The vertical dashed line separates the clusters A1291A
and A1291B. The galaxies within 30\arcmin of the cluster centers
are presented.
}
\end{figure}
\end{document}